\documentclass[10pt,twocolumn,letterpaper]{article}

\usepackage[pagenumbers]{cvpr} 
\usepackage[dvipsnames]{xcolor}

\definecolor{cvprblue}{rgb}{0.21,0.49,0.74}
\usepackage[pagebackref,breaklinks,colorlinks,citecolor=cvprblue]{hyperref}
\usepackage{enumitem}
\def\confName{CVPR}
\def\confYear{2024}

\title{Choreographing the Digital Canvas: A Machine Learning Approach to Artistic Performance}

\author{Siyuan Peng, Kate Ladenheim, Snehesh Shrestha, Cornelia Ferm\"uller\\
University of Maryland, College Park\\
College Park, Maryland\\
{\tt\small peng2000@umd.edu, klad@umd.edu, snehesh@umd.edu, fermulcm@umd.edu}
}

\begin{document}
\maketitle

\begin{abstract}
 This paper introduces the concept of a design tool for artistic performances based on attribute descriptions. To do so, we used a specific performance of falling actions. The platform integrates a novel machine-learning (ML) model with an interactive interface to generate and visualize artistic movements. Our approach's core is a cyclic Attribute-Conditioned Variational Autoencoder (AC-VAE) model developed to address the challenge of capturing and generating realistic 3D human body motions from motion capture (MoCap) data. We created a unique dataset focused on the dynamics of falling movements, characterized by a new ontology that divides motion into three distinct phases: Impact, Glitch, and Fall. The ML model's innovation lies in its ability to learn these phases separately. It is achieved by applying comprehensive data augmentation techniques and an initial pose loss function to generate natural and plausible motion. Our web-based interface provides an intuitive platform for artists to engage with this technology, offering fine-grained control over motion attributes and interactive visualization tools, including a 360-degree view and a dynamic timeline for playback manipulation. Our research paves the way for a future where technology amplifies the creative potential of human expression, making sophisticated motion generation accessible to a wider artistic community.

\end{abstract}    
\section{Introduction}
\label{sec:intro}
Even before the rapid development of AI, the living performing arts industry was seeing an increasingly closer collaboration between creative and digital tools. Nowadays, with accessible professional software, Generative AI has revolutionized other creative industries, such as image and video editing. However, the field of dancing performance still lacks AI support. With limited tools, performers and choreographers need help visualizing and materializing complex movement ideas. Current human pose generation models offer little granularity -- many of them use a single action category as auxiliary input \cite{gopalakrishnan2019neural,ormoneit2005representing,zhou20193d,zhou2021hierarchical,CombineRNNAT,li2017auto,petrovich2021action,yang2018pose} or a description \cite{zhang2022motiondiffuse, tevet2022human, lin2018human} of a single action category. This limits the nuanced controls that artists need to convey their unique vision. A dedicated AI-driven movement generation tool can unlock new dimensions in performance by offering an unprecedented range of motion exploration, enhancing creative freedom, and facilitating intricate choreography design. 
\begin{figure}[t]
  \centering
   \includegraphics[width=0.9\linewidth]{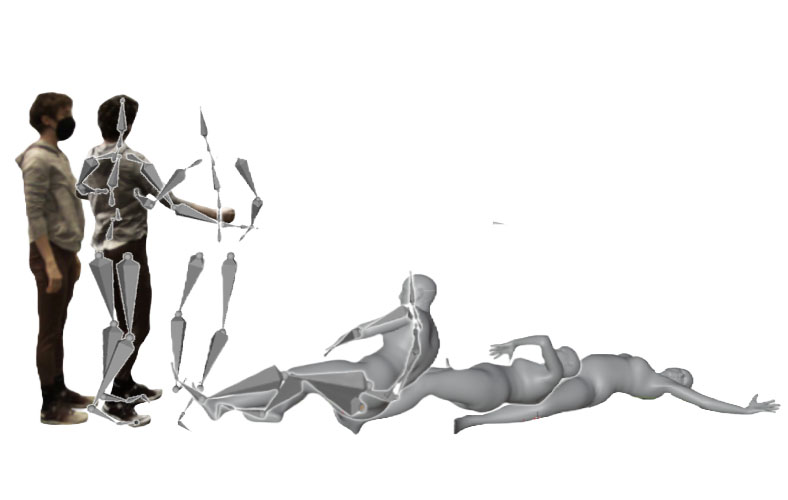}

   \caption{Dynamic motion capture sequence illustrating the transition between RGB recording to MoCap to a refined 3D animation. It demonstrates our model's comprehensive process, capturing the nuanced dynamics of human movements and translating them into realistic animated sequences.}
   \label{fig:teasing image}
\end{figure}
In this work, we aim to leverage a combination of "falling movement sequences" attributes to generate a myriad of 3D human motions that encapsulate the essence of an artist's intent. By constructing a granularly controlled, attribute-conditioned model, we give artists the ability to experiment with and explore the vast potential of movement in their creative process. 

With no existing dataset, we collected recordings of a single artist performing dramatic falls that conformed to the movement score for "Animating Death." This score is part of choreographies of falling and dying \cite{Lecture}. This project includes artistic works ``Monumental Death'' \cite{LadenheimMD} and ``COMMIT!'' \cite{LadenheimCOMMIT}, which use these dramatic falling choreographies as primary material. Falls in these works conform to the aforementioned score. During data collection, the artist performed falls based on a sequence of attributes: Impact, Glitch, and Fall. The artist performed falls with randomized attributes in the three phases to maintain parameters across the recordings. Utilizing the markerless Captury MoCap System~\cite{Captury}, we collected approximately 150 trials of the artist performing dramatic falling actions labeled with these attributes and granular sub-definitions of expressive motion.

Unlike previous works, the falling movement is complex and has multi-phase labels. Most of the existing datasets and task does not involve displacement and rotation of the human center. Thus, from a computational point of view, it is more challenging to represent and generate the falling movements accurately. We tackle this problem by using an RNN-style CVAE. Instead of generating motions frame by frame, we treat each phase or recurrent cycle as a whole and generate all the poses in one output. Moreover, we introduced whole-body movement data augmentation using Fourier transformation. 

Our contribution extends beyond the model itself. We present an innovative visualization tool that is an interactive canvas for artists. This web-based platform allows intuitive exploration of generated movements, with an interface that supports 360-degree viewing and a timeline for precise control over playback. It is a space where technology meets art, enabling performers to see, interact with, and refine the digital embodiment of their artistic expressions.

Our contributions to human pose representation and animation are as follows.
\begin{enumerate}
    \item We collected a unique falling pose dataset with multi-attribute labeling.
    \item We developed an attribute-conditioned cyclic 3D human body motion synthesis model.
    \item We built a user-friendly interactive visualization tool for generated human movements.
\end{enumerate}

Further, our work presents a unique collaboration between artists and computer scientists; one in which the point of view of a particular artist drives the creative output of a machine-learning model. Instead of animations built off of aggregate data, ultimately erasing the identities and particularities of the contributing performers, our resulting tool celebrates the particular creative vision and embodied attributes of a single artist. The resulting animation tool offers new creative possibilities for falling animations, which could be extended to various other choreographed motions.
\section{Related Work}
\label{sec:relatedwork}

\textbf{Machine Learning}: Prior to deep learning, researchers applied optimization methods to 3D human motion prediction and synthesis tasks \cite{guerra2005discovering,li2010learning}. Methods like inverse kinematics \cite{KHATIB2009211, Katsu2004Synthesizing} and motion graphs \cite{Arikan2003Motion}, however, need manual tuning and cannot generate complex and diverse human movements. With the recent development of generative models like GANs \cite{Goodfellow2020Diffusion} and Diffusion \cite{ho2020denoising}, 3D human body motion tasks have received significant attention. Yang, Ceyuan, et al. \cite{yang2018pose} utilize GANs on pose sequences and semantic consistency to control human motion dynamics. With the help of large motion datasets, Lin and Amer \cite{lin2018human} treat class labels as text conditioning and feed it to an RNN-based GAN network. \cite{Barsoum_2018_CVPR_Workshops} build a probabilistic function conditioned on previous frame actions. The limitations of GAN-based networks include accumulated errors in long sequences; it is difficult to train them and to model spatial information. Denoising Diffusion models have shown remarkable performance in generating diverse and realistic images and videos. Recent works have adopted this methodology for motion modeling with promising results. The MDM model \cite{tevet2022human} is a transformer-based diffusion model designed for various tasks, including text-to-motion and action-to-motion. A significant contribution is that it predicts on samples rather than noise. MotionDiffuse\cite{zhang2022motiondiffuse}, a diffusion-based human motion synthesis model, is capable of responding to fine-grained manipulation of body parts. PhysDiff\cite{Yuan_2023_ICCV} is designed to integrate physical constraints into the diffusion process, enhancing the physical plausibility of existing models. However, the downside of the diffusion-based model is the need for a vast amount of data to generate high-quality and diverse motion sequences. Variational Autoencoder (VAE) has been a popular method for solving human body motion synthesis tasks. Habibie et al. (2017) \cite{Habibie2017recurrent} designed a VAE with a recurrent design, showing the potential of VAEs in capturing the temporal dependencies. Yan et al. (2018) \cite{Yan_2018_ECCV} utilize the concept of motion modes to design their MT-VAE model capable of generating multiple diverse facial and full-body motions. Generating motion frame by frame, He et al. (2018) combine the VAE and RNN design to generate consistent and diverse video sequences. Our work builds upon Petrovich et al.'s ACTOR\cite{petrovich2021action} network design to extract sequence-level embeddings and generate holistic body movement.

\noindent\textbf{Artistic Animation:} Our work also builds on embodied data collection, translation, and generative animation in artistic contexts. Shaw \cite{Shaw} describes projects \textit{Synchronous Objects} and \textit{Motion Bank;} the former presenting alternative visualizations of embodied data and the latter providing a platform for annotating choreography specific to an artist's vision. Choreographic motion capture tools have been explored by Whitley \cite{Whitley} and collaborators, though this project creates sequences from prerecorded motions. Wayne McGregor's \textit{Living Archive} \cite{McGregor} uses machine learning processes to generate new choreography from the embodied data of McGregor's dancers. Ellsworth \cite{Ellsworth} used GANs in the artwork \textit{Cellular Automaton} to extend spatial configurations of pre-recorded motions. 

\noindent\textbf{Dataset:} Previous works have collected various human body movement datasets, including the popular Human3.6M dataset \cite{h36m_pami}. It contains 3.6 million human poses in 17 scenes. The UESTC dataset \cite{ji2019large} has  2.5 thousand movement sequences in 40 simple action categories. With less recorded data, the HumanAct12 dataset \cite{guo2020action2motion} provides joint coordinates for 12 action categories. 

\section{Dataset}
\noindent\textbf{Data Collection.} We used the Captury \cite{Captury} markerless motion capture system to record human motion sequences. Eight MoCap cameras in a circular formation, four at high and four at low altitudes were used to ensure good recording of the actor's poses when on the ground. For each trial, the actor's body center location, rotation, and the relative rotations of 24 bones in a kinematic tree were recorded in line with the SMPL model \cite{loper2023smpl}. Before inputting into the model, we set the starting location of the sequences to be the world origin. With the benefit of having a smooth loss function, 6D representations \cite{zhou2019continuity} of the angles were used in body center rotation and bone's relative rotations representation. 

\noindent\textbf{Attribute Description.} We organize all falls into a new ontology consisting of a five-part (impact, glitch, fall, end, and resurrection) movement score for "animating death" \cite{LadenheimMD, LadenheimCOMMIT}. 
Our collection process focused on the first three parts of the score: the Impact, the embodied site of initiation for the motion; the Glitch, a performed moment of panic, confusion, or shuddering ecstasy that sets the stage for the next segment; and the Fall, the passage of the body from upright to lying down. Our dataset specified body areas of initiation (head, torso, arms, and legs), which were impacted described by the following qualities: 
     \textbf{Push}: the impacted body part appears to be shoved in any direction;
     \textbf{Prick}: a localized, sharp action like a needle piercing skin;
     \textbf{Shot}: a forceful, localized action akin to a gunshot;
     \textbf{Contraction}: the hollowing out or concaving of an area of the body; 
     \textbf{Explosion}: a violent, bursting action starting at the point of impact and radiating quickly outwards and away from the body.

The next phase of the choreography is Glitch, which has its own set of aesthetic parameters: 
     \textbf{Shake}: a quake or tremor;
     \textbf{Flail}: wild, out-of-control motions that extend outwards in a flinging motion;
     \textbf{Flash}: a single brief, spreading motion akin to a flash of light; 
     \textbf{Stutter}: repetitive stop and start motions;
     \textbf{Contort}: twisting or warping the body; 
     \textbf{Stumble}: an off-balance, tripping quality; 
    \textbf{Spin}: turning around an axis, which can be full-bodied (the performer turns around their center line) or localized (for example, the spinning of the hand in a circle);
     \textbf{Freeze}: momentary stillness

Finally, we defined qualities for the Fall phase: 
     \textbf{Release}: where the muscles of the body relax and collapse;
     \textbf{Let Go}: the feeling of a cord being cut, or whatever is holding the body up disappears;
     \textbf{Hinge}: a reference to Horton and Graham modern dance techniques, where the body slowly descends to the ground in a flat shape akin to a door hinge; 
     \textbf{Surrender}: a performance of pleading or giving up, often accompanied by kneeling or other diminutive postures;
    \textbf{Suspend}: for those familiar with Laban efforts, suspension has a float effort. A slow passage to the ground with slightly upward oppositional energy; suspensions are lightly supported and less angular than a hinge.

\section{Method}
\label{method}

\begin{figure*}[t]
  \centering
   \includegraphics[width=0.8\linewidth]{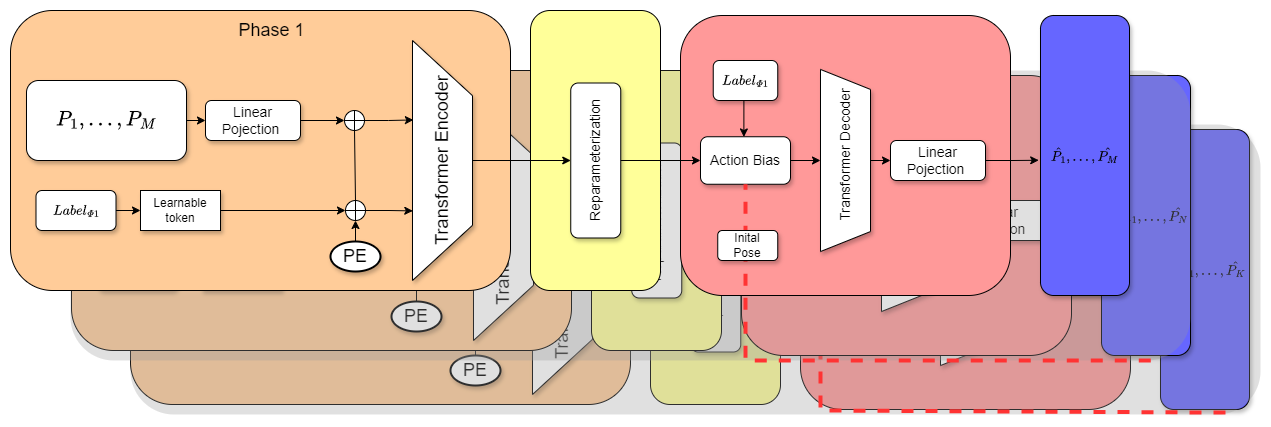}

   \caption{\textbf{Attribute-conditioned Variational AUtoencoder model architecture}: Taking a sequence of body poses, we split them into inputs for different phases. The model learns to reproduce the movement specifically for each phase. The final multi-phase sequences are the concatenation of all outputs from the three phases. Improved from the ACTOR model \cite{petrovich2021action}, we employed a cyclic design (indicated by the dotted red line) where the last frame from the previous phase serves as the initial guidance of the generation.}
   \label{fig:model_architecture}
\end{figure*}

\subsection{Problem setup}
When describing human actions, it's desirable to separate body shape from pose \cite{osman2020star, pavlakos2019expressive,xu2020ghum}. Ignoring body shape, we aim to generate a sequence of pose parameters, specifically, the relative joint rotation in the kinematic tree of the human body. In formal notation, given a combination of falling attribute labels $L_\phi{1}, L_\phi{2}, L_\phi{3}$ (phase $\phi$), and time intervals ${1,...N}$, we synthesize a sequence of body joint parameters $\hat{P_1} ... \hat{P_N}$, that contain the root joint's rotation and translation, as well as the body joint rotation parameters as shown in Fig \ref{fig:model_architecture}.

\noindent\textbf{Data Augmentation.} Due to the limited amount of data and our aim of building a more robust model, we performed data augmentations on existing data. We began by applying a Fast Fourier Transform (FFT) to convert segments of motion time-series data into the frequency domain. This transformation allows us to manipulate the motion data in ways that are not easily achievable in the time domain. We then tweaked the magnitude and phase information and converted it back to the time domain, ensuring the augmented motion remains continuous and retains the natural flow of human movement. We noticed our model struggled to create movements from specific starting poses. This issue was linked to the limited training data in which the performer mostly started from the same direction. To fix this, we added variations by randomly rotating the data around the human body's vertical axis. This approach helped our model learn to generate movements from various starting positions, making it more versatile.

\subsection{CVAE Model}
We build on the CVAE-based ACTOR model \cite{petrovich2021action} but use action attributes as conditions and extend the model with an RNN-style design. Specifically, we employ separate encoder and decoder pairs for each of the three distinct action phases, facilitating precise extraction and generation of poses from latent spaces. The last frame of the generated video is used as the initial guiding pose for the next phase. This design choice contributes to the accurate representation of each phase. Our model generates more realistic and accurate human falling movements than a single-phase autoencoder. This will be explained in detail in the result section. Despite the model's cyclical structure, we adopt a concurrent training strategy for all three phases using the ground truth data from the MoCap Dataset. However, we need to generate poses phase by phase during inference time. 

\noindent\textbf{Encoder.} The input are a sequence of body poses from $P_1$ to $P_K$, where $P_1$ to $P_M$ are poses for phase 1, $P_{M+1}$ to $P_N$ are inputs for phase 2, and $P_{N+1}$ to $P_K$ are for phase 3. Since the operations are similar for each phase, we will only explain the steps in detail for phase 1. After converting the phase label into learnable tokens, we prepend them to the pose sequences. These tokens, similar to those used in \cite{petrovich2021action}, are used for pooling purposes in the temporal dimension. Due to the self-attention mechanism, these tokens will aggregate (or pool) information from the entire action sequences. A similar implementation can be seen in the BERT \cite{devlin2018bert} model for sentiment prediction. Positional Encoding (PE) has been proven to be a vital part of various works, such as Transformer \cite{vaswani2017attention} and NeRF \cite{nerf} architectures. We also take advantage of it and add it to the input, and the encoder encodes all data into a low-dimensional latent space. To extract the distribution parameters $\mu$ and $\Sigma$, we take the first two outputs of the corresponding encoder for each phase.

\noindent\textbf{Embedded Space.} After extracting the two distribution parameters $\mu$ and $\Sigma$ from the encoder output, we use the reparameterization trick introduced in \cite{kingma2013auto} to allow gradients to pass in the sampling process. Specifically, instead of sampling the latent vector $z$ directly from the distribution, $z$ is sampled using equation \ref{equ:reparm}, where $\otimes$ represents element-wise multiplication and $\epsilon \sim \mathcal{N}(0,1)$. 
\begin{equation}
\label{equ:reparm}
    z = \mu + \sigma \otimes \epsilon
\end{equation}

\noindent\textbf{Decoder.} The goal of the decoder is, given a latent vector $z$ and one of the falling attributes $L_{a_i}, L_{b_j}, L_{c_k}$, to generate a novel sequence of human body parameters $\hat{P_1} ... \hat{P_N}$. Given the structural similarity among our three attention decoders, we detail the operation of a single decoder as a representative example. To add falling attributes to the decoder, we sum a learnable token with the latent space to shift it to an attribute-dependent space. Moreover, different from the previous decoder implementation \cite{petrovich2021action}, we have the initial pose from the previous phase as an auxiliary input. Combining the initial pose and given attribute, we add them and feed them to the attention decoder. The output of the decoder will be passed into a fully connected layer and generate the entire motion sequences for the phase: $\hat{P_1}$ to $\hat{P_M}$. After all three decoders synthesize the outputs, we concatenate all generated poses.

\noindent\textbf{Loss.} We utilize a combined loss comprising the human body model's parameters, KL divergence, vertex reconstruction loss, and initial pose loss. The human body model's parameter loss is an L2 loss on the human model center's location and rotation and the relative bone rotation with respect to the parent bone. KL divergence loss measures the divergence between the predicted and actual distributions of the latent variables, helping to regularize the model and ensure a smooth latent space. Vertex reconstruction loss is the L2 difference between the vertexes of the reconstructed SMPL human model. This loss ensures the geometric fidelity of the reconstructed human model to the real human body shapes and postures. An additional L2 loss is applied to the initial pose's model parameters and reconstructed vertices. This loss is specifically designed to enhance the accuracy of the generated sequence's starting pose, ensuring it aligns closely with the targeted initial conditions.

\subsection{Interactive Interface}
\begin{figure*}[t]
  \centering
   \includegraphics[width=0.8\linewidth]{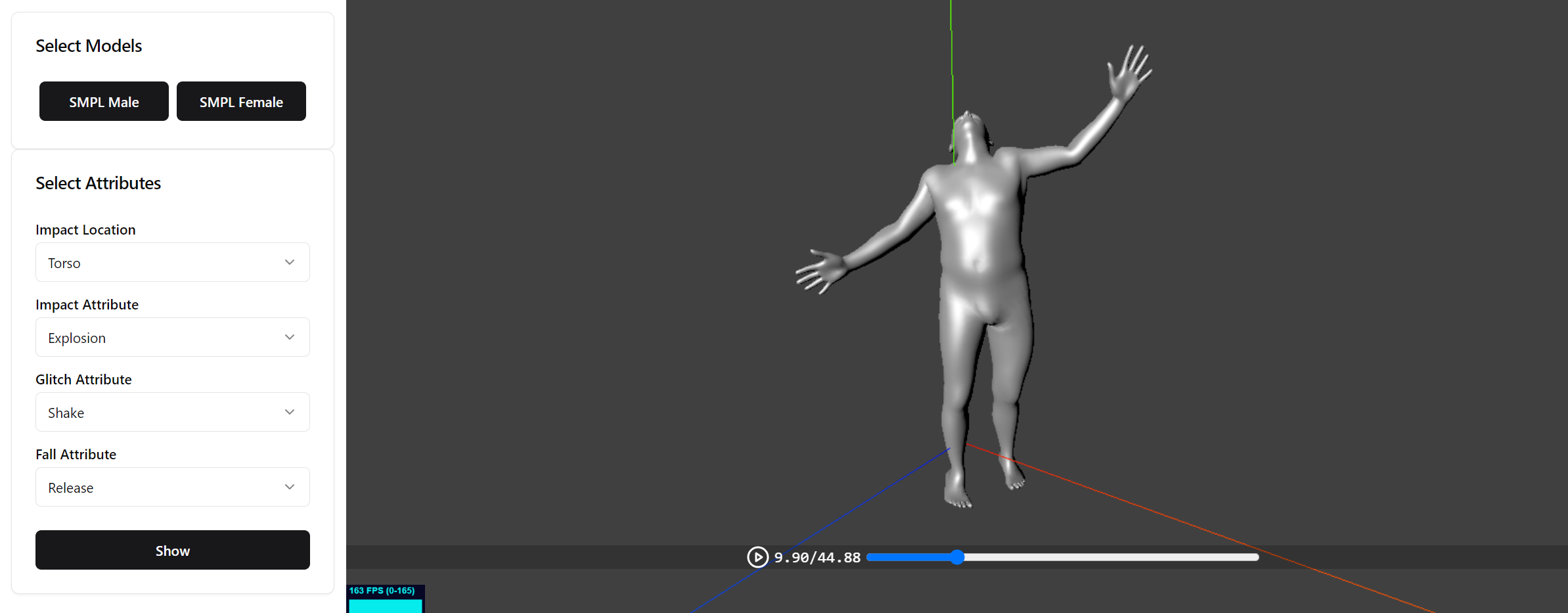}

   \caption{Interface of the 3D human motion visualization tool showcasing the model and attribute configuration selection menu. Users can select between SMPL male and female models and customize the motion attribute for impact location, impact attribute, glitch attribute, and fall attribute. The interactive 3D viewer on the right displays the generated pose according to the selected attributes, with a slider for navigating the motion timeline. }
   \label{fig:interface}
\end{figure*}
A nexus between the performer's artistic view and AI-generated movements, the interactive interface is an intuitive web-based platform that allows users to manipulate 3D motions with ease. On the left side of the interface is the menu where users can select a range of motion attributes to synthesize falling movement sequences. Taking advantage of the SMPL model, we can also switch the animation human model between female and male, ensuring a broad application across diverse artistic contexts. This adaptability empowers choreographers to visualize the same sequence performed by different genders, offering valuable insights into the interplay between movement and the performer’s physicality. On the right side is the 3D animation of the poses. It offers a 360-degree view, allowing every subtle inflection of movement to be examined from any angle. Moreover, a responsive timeline gives the user control over the playback, making it possible to pause, play, and scrub through the sequence to focus on moments of interest. This level of interactivity enables the performer to delve into each artistic choice -- adjusting the parameters on the fly and viewing the results immediately. 

\noindent The interface we have developed while adopting the intrinsic dynamic of the falling movement is by no means limited to one artistic performance. It's a versatile platform designed to accommodate various artistic expressions. Whether it is the elegant gestures of ballet, the cultural reflection of folk dances, or the abstract expressions found in modern dance, our platform is equipped to handle the generation and visualization of these diverse performances.

 \section{Results}
\subsection{Experiment Setup}
To improve the robustness of our model, we initiated our training process by incorporating pre-trained weights of the ACTOR model\cite{petrovich2021action}, which is trained on the UESTC dataset\cite{ji2019large}, and applying these weights across all encoder-decoder branches of our architecture. We selectively retained weights compatible with our model's framework and discarded unfitted ones. We trained our model on the falling dataset for 2000 epochs, using a batch size of 20 and a learning rate of 1e-4.

\subsection{Quantitative  Result}
 Our approach to evaluating the proposed model aligns with the metrics previously established in the literature \cite{guo2020action2motion, petrovich2021action}, specifically focusing on action recognition accuracy, Fréchet Inception Distance (FID) score, and diversity. The choice of these metrics allows for a comprehensive analysis of the model's performance in generating realistic and varied motion sequences that align accurately with the intended actions. To effectively capture these measurements, a customized recognition model, modified based on STGCN \cite{yan2018spatial}, is used for extracting the feature embedding and classification. Instead of having one classification head, our recognition model implements a multi-branch classification. Each branch evaluates one specific action phase, ensuring a nuanced assessment of the model's performance across different motion stages. Feature embeddings are extracted immediately preceding the multi-branch classification heads. These embeddings encapsulate the high-dimensional characteristics of the action sequences, serving as the foundation for accurate classification and the calculation of the FID score. 

 The results, as detailed in the table \ref{tab:quantitative}, demonstrate that our models have superior performance in terms of the accuracy and the diversity of the generated actions. Although our model is not as good as the ACTOR regarding the FID score, we reason that while it may prioritize accuracy and diversity, it might capture some aspects of the action dynamics that are not fully represented in the real data used to compute the FID. 
\begin{table}
  \centering
  \begin{tabular}{@{}lccc@{}}
    \toprule
    Method & FID $\downarrow$ & Accuracy $\uparrow$ & Diversity $\uparrow$\\
    \midrule
    Ours (Addition) & 195.23 & 0.289 & 11.95\\
    Ours (Concate) &  176.50 & 0.288 & 12.31\\
    ACTOR & 118.28 & 0.251 & 10.20 \\
    \bottomrule
  \end{tabular}
  \caption{Comparative Evaluation of our model and the ACTOR model using Frechet Inception Distance (FID), recognition accuracy, and generation diversity metrics. The table presents the performance of two variants of our model either with Addition or  Concatenation -- alongside the ACTOR model. A lower FID value indicates better performance. Higher accuracy and higher diversity values are better. This result demonstrates that while our models show higher diversity and comparable accuracy, the ACTOR model achieves a lower FID score.}
  \label{tab:quantitative}
\end{table}

\subsection{Qualitative Result}
In assessing the qualitative aspect of the motion generation, we generated three sets of around 500 motions with random attributes using our RNN-style autoencoder and ACTOR model \cite{petrovich2021action}. Evaluated by a group of performance and art professionals in a double bind setting, the generated poses are assessed in three critical dimensions: accuracy of the generation, plausibility of the poses, and the smoothness of the movement. All reviewers rate on a Likert scale from 1 to 10, with 1 being the lowest and 10 the highest possible score. These scores reflect the perceived quality and realism of the generated motions. Based on the results as shown in table \ref{tab:qualitative}, our model outperforms the ACTOR model in generating human poses that are more accurate, plausible, and smooth. 

\begin{figure*}[h]
    \centering
    \begin{minipage}[b]{\textwidth}
        \includegraphics[width=\textwidth]
            {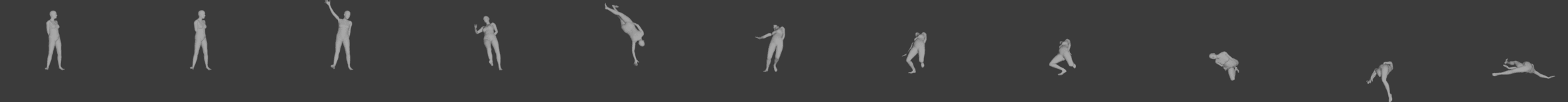}
            \caption{Impact Location: arms; Impact Attribute: prick; Glitch Attribute: flail; Fall Attribute: release}
            \label{fig:result_1}
    \end{minipage}
    \begin{minipage}[b]{\textwidth}
        \includegraphics[width=\textwidth]
            {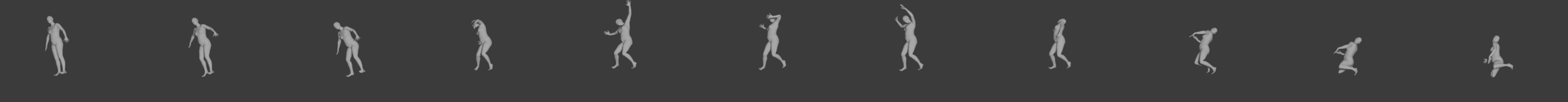}
            \caption{Impact Location: torso; Impact Attribute: push; Glitch Attribute: contort; Fall Attribute: surrender}
            \label{fig:result_2}
    \end{minipage}
    \begin{minipage}[b]{\textwidth}
        \includegraphics[width=\textwidth]
            {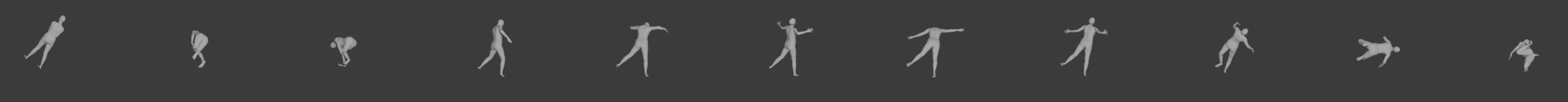}
            \caption{Impact Location: torso; Impact Attribute: contraction; Glitch Attribute: spin; Fall Attribute: let go}
            \label{fig:result_3}
    \end{minipage}
    \begin{minipage}[b]{\textwidth}
        \includegraphics[width=\textwidth]
            {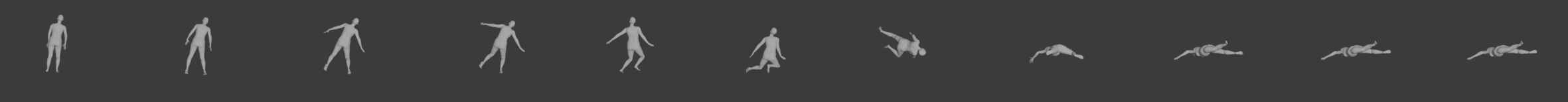}
            \caption{Impact Location: head; Impact Attribute: shot; Glitch Attribute: stumble; Fall Attribute: release}
            \label{fig:result_4}
    \end{minipage}
    \begin{minipage}[b]{\textwidth}
        \includegraphics[width=\textwidth]
            {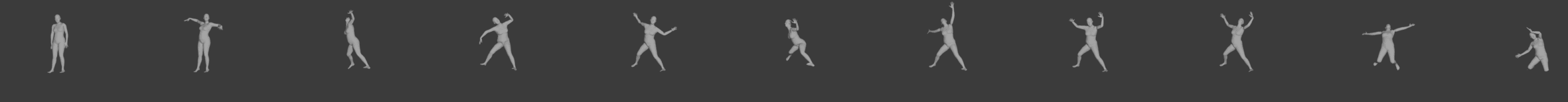}
            \caption{Impact Location: arms; Impact Attribute: explosion; Glitch Attribute: stutter; Fall Attribute: surrender}
            \label{fig:result_5}
    \end{minipage}
    \begin{minipage}[b]{\textwidth}
        \includegraphics[width=\textwidth]
            {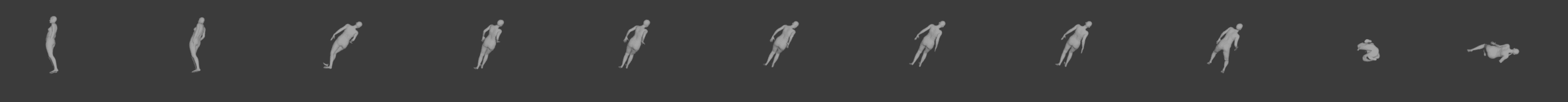}
            \caption{Impact Location: torso; Impact Attribute: push; Glitch Attribute: freeze; Fall Attribute: let go}
            \label{fig:result_6}
    \end{minipage}
    \begin{minipage}[b]{\textwidth}
        \includegraphics[width=\textwidth]
            {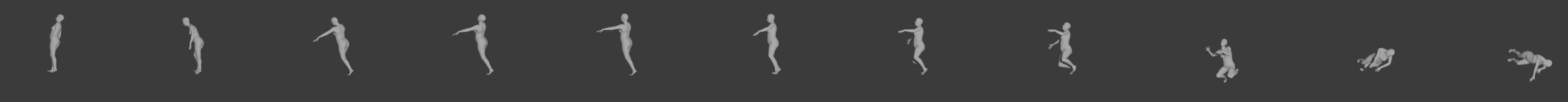}
            \caption{Impact Location: torso; Impact Attribute: explosion; Glitch Attribute: shake; Fall Attribute: hinge}
            \label{fig:result_7}
    \end{minipage}
\end{figure*}

From fig \ref{fig:result_1} to fig \ref{fig:result_7}, we present a selection of generated samples from the model. They illustrate a diverse range of motions that can be synthesized when varying the attributes within the model.
\begin{table}
  \centering
  \begin{tabular}{@{}lccc@{}}
    \toprule
    Method & Accuracy &Plausibility & Smoothness \\
    \midrule
    Ours (Addition) & 5.130 & 4.841 & 4.548 \\
    Ours (Concate) & 6.117 & 5.464 & 5.120 \\
    ACTOR & 5.148 & 4.791 & 4.509 \\
    \bottomrule
  \end{tabular}
  \caption{Comparative Evaluation of our models and the ACTOR model rated by arts professionals. This table shows the ratings of attribute accuracy, movement physical plausibility, and motion smoothness given by the artists. Each model was assessed on a Likert scale, with higher scores indicating better performance in each category. The results highlight that our Concatenate variant is rated favorably in all three categories, suggesting it more effectively captures artistic movements.}
  \label{tab:qualitative}
\end{table}

\subsection{Ablation Study}
We examine two distinct methods for combining features before they are fed into the decoder: concatenation and addition. Specifically, we need to integrate the embedded initial pose with the attribute-dependent feature space vector $Z$. This exploration aims to understand how different methods of combining these features impact the effectiveness of our model.

\subsection{Limitation}
We implemented data augmentation by varying the amplitude of the movement. However, due to the randomness of the action and the augmentation process, there were instances where parts of the human body, such as the elbow, ended up inside the human model. Future work could explore more realistic and robust methods of data augmentation. The training time and resource requirements increased as we implemented an RNN-style design to the model. This is compounded by the fact that we input entire motion sequences into the model. Longer input motions would further escalate the need for computational resources.

\section{Conclusion}
We presented the concept of a novel design tool for artistic performances described by attributes and implemented it for one performance. We collected and curated a unique dataset featuring complex labeling, and developed a novel attribute-conditioned variational autoencoder designed to focus on different phases of the input data. The cyclic design achieved more granular control over the motion sequences. Despite the limited amount of data, we successfully generated diverse and realistic 3D human movements by applying data augmentation and initial pose loss.
{
    \small
    \bibliographystyle{ieeenat_fullname}
    \bibliography{main}
}


\end{document}